\documentclass{ws-procs975x65}

\def\apj{{\em ApJ}}
\def\apjs{{\em ApJS}}
\def\aap{{\em A\&A}}
\def\aj{{\em AJ}}
\def\mnras{{\em MNRAS}}
\def\nat{{\em Nature}}
\def\apjl{{\em ApJL}}

\def\pasp{{\em PASP}}

\def\etal{{\it et al.\, }}

\begin{document}

\title{Extragalactic jets: the high energy view}

\author{F. Tavecchio}

\address{Osservatorio Astronomico di Brera-Merate, via Bianchi 46, 23807
Merate, Italy\\E-mail: tavecchio@merate.mi.astro.it}




\maketitle

\abstracts{I review the current knowledge of high-energy emission from
extragalctic jets. First I discuss $\gamma $-ray emission from
blazars, which provides us numerous precious information on the
innermost portions of the relativistic jets. I describe the
constraints on the dynamics of the jet from the subpc to the pc scale
provided by recent VLBI studies of TeV sources, together with the
modelling of the emission from the blazar jet. Finally I discuss high
energy emission from large scale jets as seen by {\it Chandra} and I
report on the expected gamma-ray emission from large-scale regions of
jets.}

\section{Introduction}

The study of relativistic jets has been revitalized in the past 15
years by the discovery of their high-energy emission, which in some
cases can extend up to the (multi-)TeV band. EGRET onboard CGRO first
discovered (e.g. Von Montigny 1995) that {\it blazars}, which form the
most active class of AGNs, are bright emitters of $\gamma $-rays in
the MeV-GeV domain. The discovery of such an intense flux of
$\gamma$-rays definitively confirmed the twenty years old proposal
(Blandford \& Rees 1978) that the emission from blazars, dominated by
a smooth non-thermal continuum from radio to $\gamma $-rays, is
produced within a relativistic jet, pointing close to the line of
sight. Indeed, in order to avoid photon-photon absorption (producing
electron-positron pairs), the compactness of the source must be small,
condition that directly implies that the emission is beamed
(e.g. Maraschi, Ghisellini \& Celotti 1992), with values of the
Doppler beaming factor $\delta=[\Gamma (1-\beta \cos \theta)]^{-1}$
(where $\Gamma$ is the bulk Lorentz factor of the flow, $\beta =v/c$
and $\theta $ is the viewing angle) larger than 5-10.  The subsequent
discovery (possible thanks to the advent of atmosferic Cherenkov
telescopes) of bright, strongly variable TeV emission from a handful
of blazars (see the recent reviews by Krawczynski 2003 and Costamante
2003), confirmed and strengthened these evidences and initiated
renewed efforts devoted to enlarge the number of known TeV sources,
especially those located at relatively large distance ($z>0.1$), since
they can be used to probe the poor known infrared background in the
1-20 $\mu$m range (e.g. De Jager \& Stecker 2002). A recent census
reports 6 sources with a firm detection in the TeV band, but in the
very near future the TeV family is expected to be greatly enlarged by
the already operative new generation of Cherenkov telescopes
(e.g. Horan \& Weekes 2003).

More recently, the discovery made by {\it Chandra} of intense X-ray
emission from large scale jets (Chartas et al. 2000; Harris \&
Krawczynski 2001), solicited new intense theoretical work to explain
the emission mechanism and the physical condition at these
scales. {\it Chandra} is now routinely discovering many new X-ray
jets, both of low-power (FRI; e.g. Worrall et al. 2001) and of
high-power (FRII; e.g. Sambruna et al. 2002, Siemiginowska et
al. 2002). Most of the models developed to understand the X-ray
emission from large-scale jets lead to predict a more or less bright
emission in the $\gamma $-ray domain: also this kind of sources is
then expected to be detected by the next-generation of high-energy
instruments, such as GLAST and the new Cherenkov telescopes.

In the following I give a short review of these topics, with
particular attention to the possibilities offered by the upcoming new
generation of high-energy instruments. For a more general review see
e.g. Maraschi (2003).

\section{Blazars}

The smooth non-thermal continuum of blazars (e.g. Urry \& Padovani
1995), extending from radio to $\gamma $-rays, is similar in both Flat
Spectrum Radio Quasars (showing luminous, QSO-like emission lines) and
BL Lac objects (with weak or even absent emission lines),
characterized in the $\nu F(\nu )$ representation by a typical
``double hamped'' shape (e.g. Fossati et al. 1998). Only after the
advent of the $\gamma $-ray telescope EGRET (onboard the {\it Compton
  Gamma-ray Observatory}) about ten years ago (Fichtel et al. 1994,
Mukherjee et al. 1997) it was possible for the first time to describe
the whole SED of blazars over the entire electromagnetic spectrum
(e.g. Von Montigny 1995).  Gamma-ray observations showed that, in many
cases, the total energetic output of blazars is largely dominated by
the high energy component; it was therefore immediately clear that
previous studies, based only on observations from the radio to the
UV-X-ray bands, lost a determinant part of the blazar phenomenology
(and power output). The extremely short variability timescales
revealed by EGRET (down to few days) confirmed that the radiation is
produced in very compact regions, and provided a strong evidence of
the presence of relativistic motion in jets (Von Montigny et al. 1995,
Dondi \& Ghisellini 1995).

The high degree of polarization (up to 20\%) convincely indicates that
the low energy component is produced through synchrotron emission.
{\it Leptonic models} assume that high energy photons are produced
through IC scattering of soft photons by the same electron population
(or pairs) responsible for the synchrotron emission.
In the Synchrotron-Self Compton model (Maraschi et al. 1992; Tavecchio
et al. 1998) it is assumed that the soft photon energy density is
dominated by the synchrotron photons themselves, while the so-called
External Compton models (Dermer \& Schlickeiser 1993, Sikora et
al. 1994, Sikora et al. 2002) assume that soft photons coming from the
external environment (accretion disk, BLR, molecular torus) dominate
over the synchrotron contribution. A sort of ``mixed'' case is
considered in the so-called mirror model (Ghisellini and Madau 1996),
in which the dominant role is played by the jet synchrotron radiation
coming back into the jet after ``reflection'' by clouds and/or free
electrons in the BLR. {\it Hadronic models}, instead, assume that the
high-energy radiation comes from reaction involving highly energetic
protons, producing $\gamma $-rays either through hadron-hadron
collisions or photon-hadron reactions, producing pairs and the
subsequent development of a e$^+$e$^-$ cascade (Mannheim 1993, Mucke
et al. 2003). Another alternative (Aharonian 2000) is offered by
synchrotron radiation from high-energy protons, coaccelerated with the
electrons responsible for the low energy component. In all the
different versions of hadronic models the magnetic field must be
rather large ($B=50-100$ G). The versions of hadronic models involving
cascade have been criticized (e.g. Sikora et al. 2002) because of the
difficulty of producing the quite flat X-ray spectra observed in FSRQ
(e.g. Tavecchio et al. 2000a). Another possibly important problem can
be identified in the fact that the pair cascade can easily saturate,
resulting in a net ``degradation'' of $\gamma $-rays into soft X-rays
(e.g. Ghisellini 2003). Therefore, although actractive because of the
possibility to also account for the acceleration of UHE cosmic rays,
hadronic models seem to face severe problems.  Note, however, that
even in leptonic models the ultimate source of power must be supplied
by protons, since the inferred power content of the electronic (and
magnetic) component alone is not sufficient to sustain the total
luminosity of the jet, that should be carried by protons (or by a
dominant magnetic field, see also below).

\begin{figure}[t]
\epsfxsize=15pc 
\epsfbox{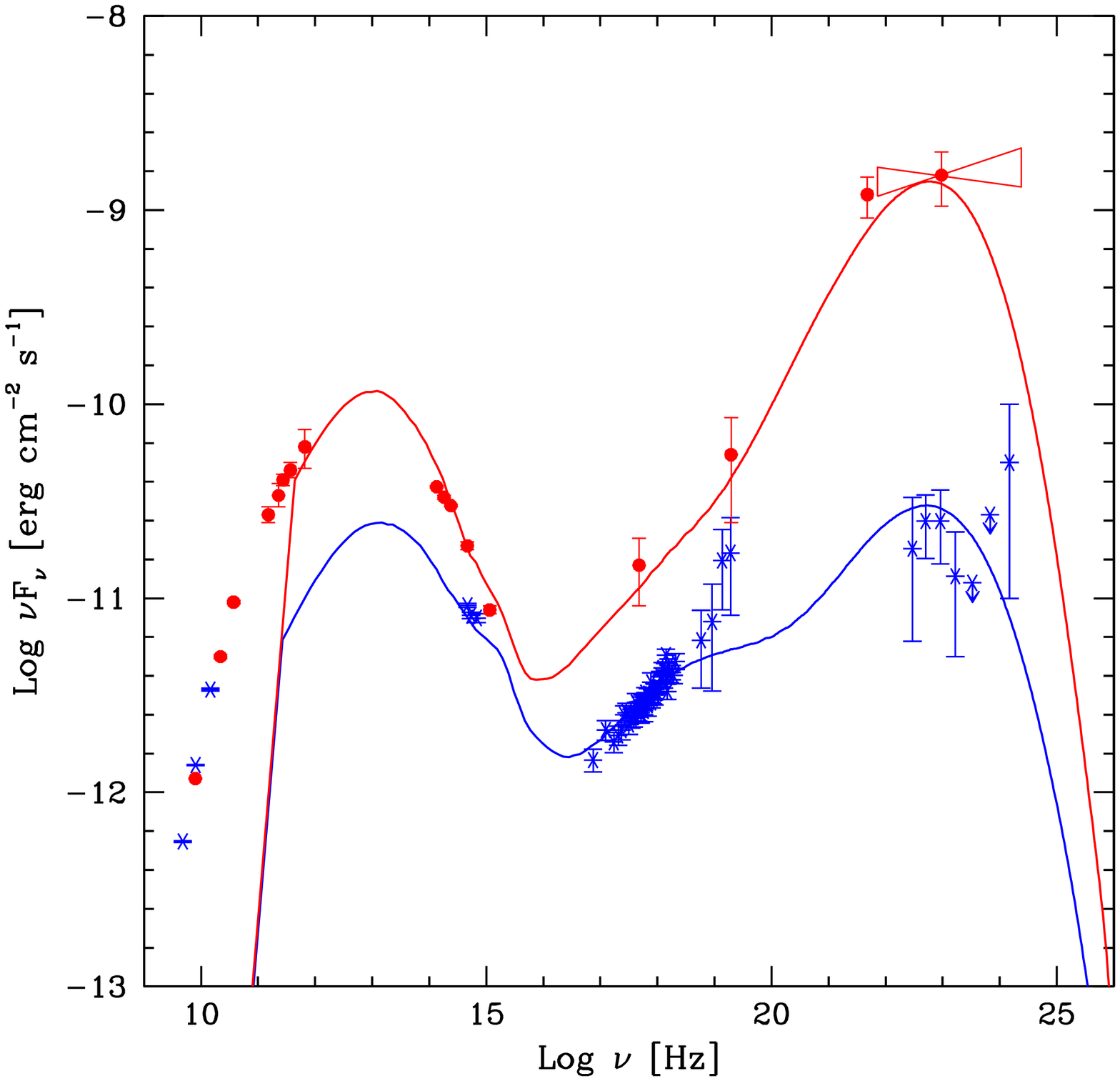}\epsfxsize=16pc\epsfbox{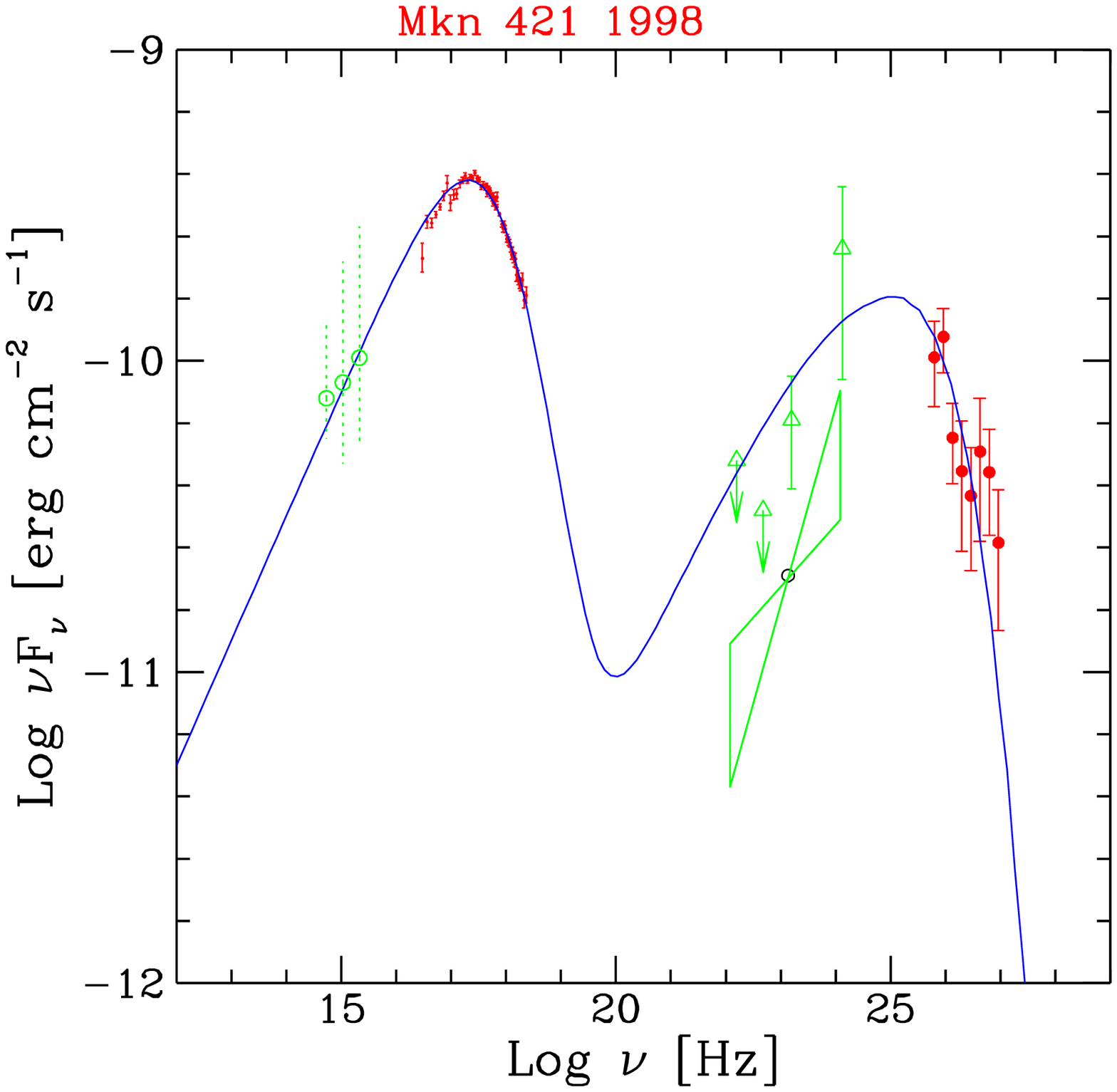} 
\caption{{\it Left:} Quasi-simultaneous SEDs of the quasar 3C279
obtained at two epochs 1996 and 1997 (from Ballo et al. 2002). The
continuous lines represent synchrotron plus Inverse Compton (including
both SSC and external radiation) models. Variability is modelled
assuming that only the bulk Lorentz factor of the flow is
changes. {\it Right:} Simultaneous X-ray and TeV spectra measured
during a flare detected during the {\it Beppo}SAX/Whipple campaign of
1998 (Maraschi et al. 1999). The solid curve is the SED obtained with
a homogeneous synchrotron/SSC model.}
\label{279}
\end{figure}

An impressive example of the phenomenology of blazars is offered by
3C279, one of the brightest sources in the $\gamma $-ray sky (it was
the first blazar discovered by EGRET, Hartman~et~al.~1992) and one of
the blazars with the best known SED.  In Fig.(\ref{279}) we report the
SED obtained during a multiwavelength campaign taken in 1996 (Wehrle
et al. 1998) and 1997. During the 1996 observations 3C279 underwent to
a large flare: in less than a week the $\gamma $-ray flux increased by
one order of magnitude. Simultaneous flares were observed in X-rays,
while at lower frequencies the flux appeared much less variable.  The
data clearly shows the double peaked SED: the synchrotron peak falls
around $10^{12}-10^{13}$ Hz, but the poor knowledge of the IR spectrum
does not allow us to determine it with precision (moreover at these
wavelength there could be a substantial contribution from the dust of
the molecular torus surrounding the central BH, see Haas et al. 1998).
The highly variable $\gamma $-ray emission (with an amplitude of about
2 order of magnitude from 1991 to the flaring state observed in 1996)
peaks just in the EGRET band, at about 1 GeV.  The overall spectrum of
3C279 can be successfully reproduced with synchrotron-Inverse Compton
emission from a single region (e.g. Hartman et al. 2001, Ballo et
al. 2002), including both internal and external seed photons.External
photons produced in the BLR could easily dominate the radiation energy
density and produce the bulk of the $\gamma $-ray emission, while SSC
could give a significant contribution in the (soft) X-ray band.

At the opposite extreme in power, Mkn 421 (Fig.1) is the brightest BL Lac
object at X-ray and UV wavelengths and the first extragalactic source
discovered (by Whipple) at TeV energies (Punch et al. 1992). It is a
typical representative of the Highly-peaked BL Lac objects: the quite
variable synchrotron component, extending from radio to X-rays, peaks in
the soft-medium X-ray range (from 0.1 to 10 keV). The position of the IC
bump is less clear, but presumably is around 100 GeV, as indicated by the
flat ($\alpha =0.6 \pm 0.1$; Hartman et al. 1999) spectrum measured by
EGRET and the steep (photon index $> 2$) TeV spectrum (e.g. Aharonian et
al. 2002, 2003a). The high-energy component is likely due to SSC, since
the optical spectrum shows almost no evidence of thermal components and
thus synchrotron photons easily dominate over any other component. A
support for the SSC model is the evidence that X-ray and TeV variability
are strongly correlated (see below), even if recent observations of
``orphane'' TeV flares are starting to make the situation more complex.

\subsection{The blazar sequence}

\begin{figure}[t]
\epsfxsize=15pc 
\epsfbox{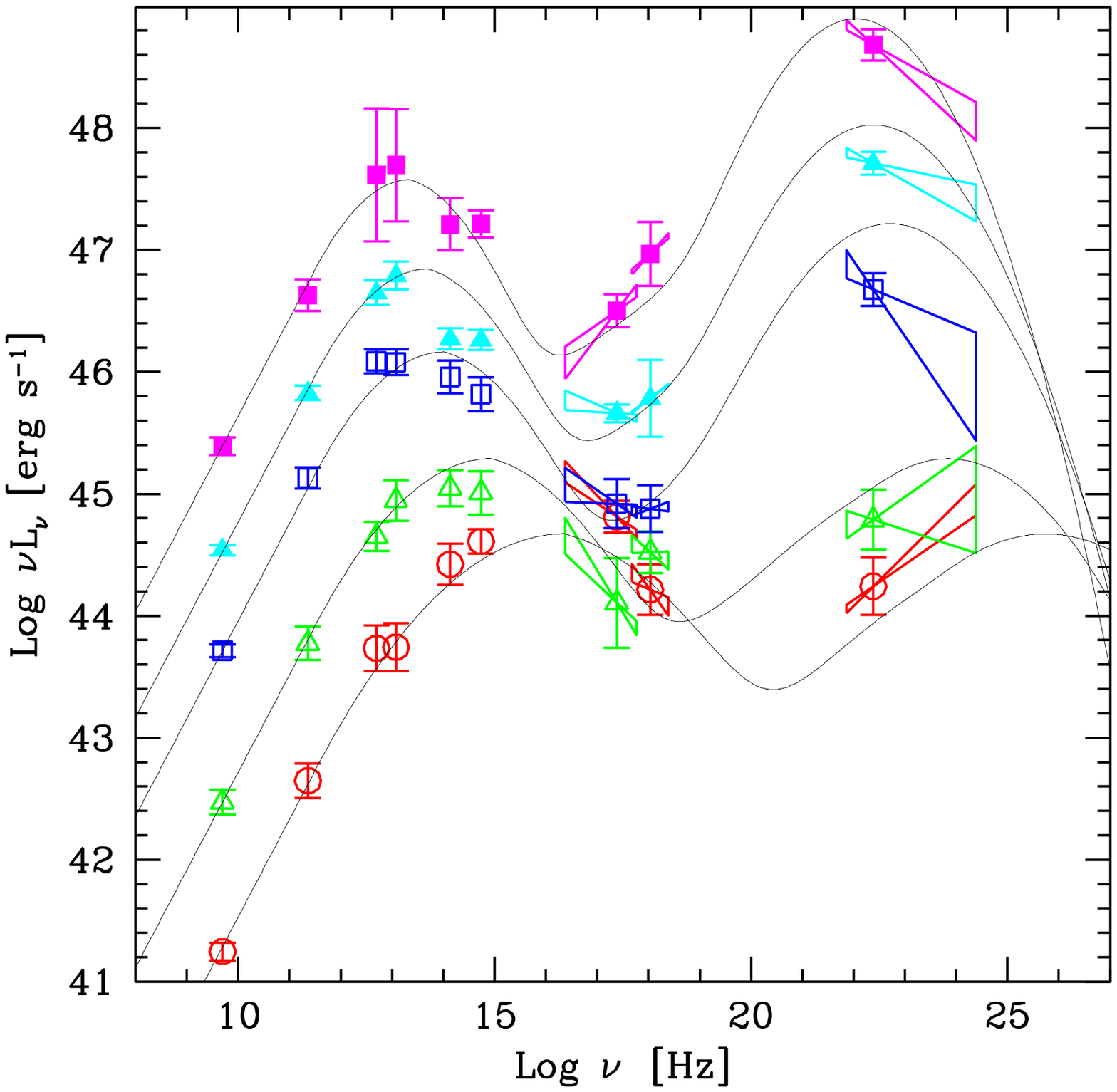}\epsfxsize=15pc\epsfbox{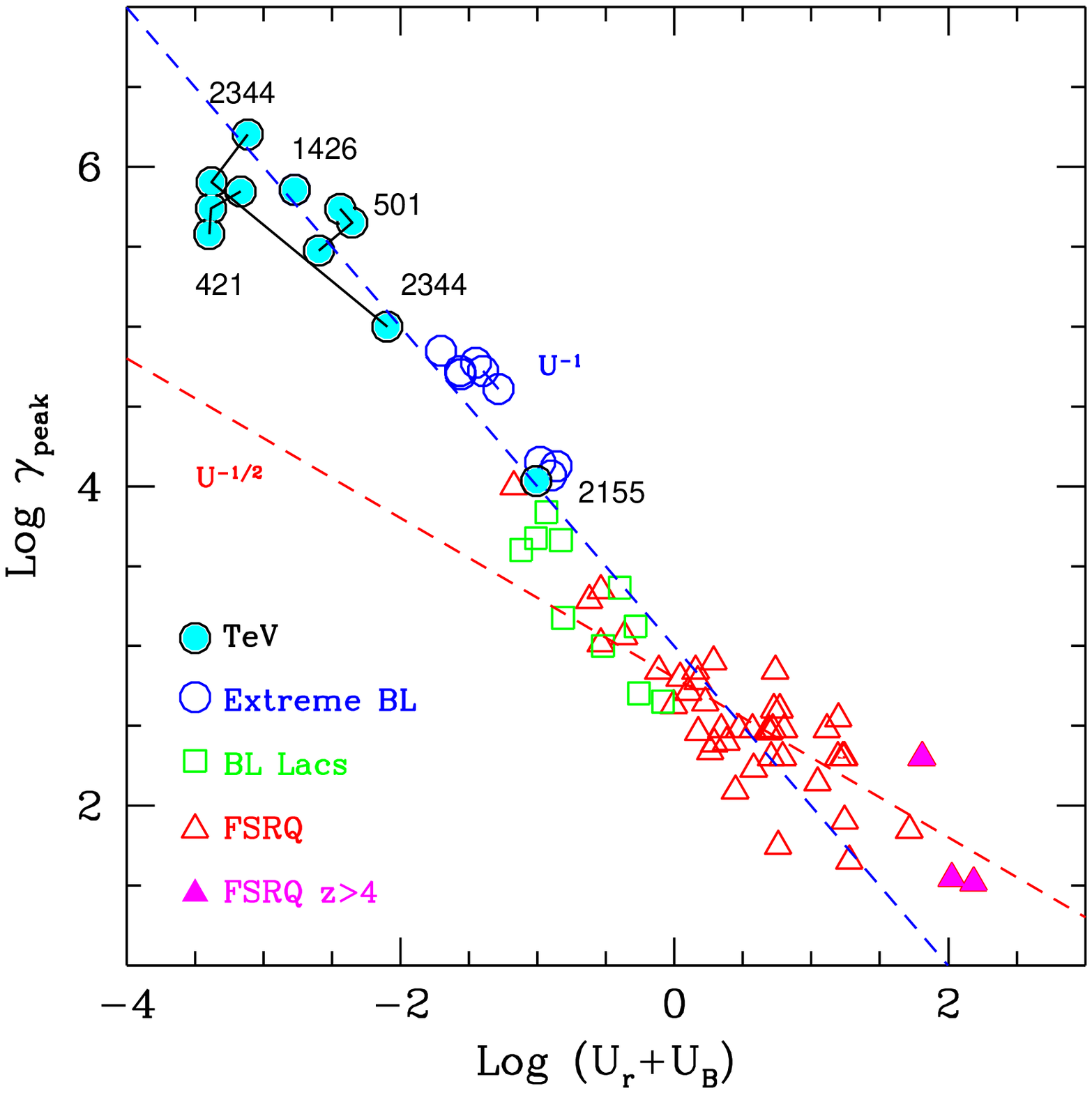} 
\caption{{\it Right:} The blazar sequence (from Fossati et al. 1998;
Donato et al. 2001).{\it Left:} The correlation (taken from Ghisellini
et al. 2002) between the Lorentz factor of particles emitting at the
peak ($\gamma _b$) and the energy density in radiation and magnetic
field. See text for a discussion.
\label{donato}}
\end{figure}

After the discovery of the high-energy component, the first systematic
exploration of the properties of the blazar SEDs up to gamma-ray
energies was done by Fossati et al. (1998).  Their procedure was to
construct ``average'' SEDs of known complete samples of blazars. Their
results led to define the so-called {\it blazar sequence}, shown in
Fig.(\ref{donato}).  They constructed average SEDs binning the objects
according to their radio luminosity, irrespective of their
classifications. Indeed, the resulting SEDs appear homogenous and
systematic trends are evident. For each luminosity class the derived
SEDs show two very broad components peaking between $10^{13} -10^{17}$
Hz and between $10^{21} -10^{24}$Hz respectively. It is evident that
sources with high luminosity output ($L\sim 10^{48}$ erg cm$^{-2}$
s$^{-1}$) have both peaks located at low energies, the first in the IR
region and the high energy peak around 1 MeV. On the contrary, at low
power the sources display very high peak frequencies: in HBL, which
appear to be the less powerful objects (with typical luminosities
$10^{42}$ erg cm$^{-2}$ s$^{-1}$) the first peak falls in the UV-soft
X-ray band and the high energy component can peak near the TeV region.
Superimposed to the averaged SEDs are curved obtained with a simple
analytic parameterization of the SED assuming that: i) that the peak
frequencies are inversely related to the radio-luminosity.  ii) that
the ratio of the two peak frequencies in each SED is constant iii)
that the height of the second peak is proportional to the radio
luminosity.  These ``analytic'' laws seem to represent the average
SEDs quite closely.


It is still an open issue whether the $\gamma$-ray properties of the
average SEDs are representative of the whole population or are
significantly biased. Clearly, since most sources are close to the
detection limit and many are known to be strongly variable
(e.g. Mukherjee et al. 1997), EGRET preferentially detected those that
were in an active state. 
There can be little doubt that the average gamma ray fluxes in
Fig.(\ref{donato}) are overestimated. Future MeV-GeV missions (AGILE,
GLAST), able to detect sources at low flux level, will help us to
better characterize the variability at high energy and to clarify
these issues.

There is a growing debate around the validity of the spectral
sequence. Despite the claims of the existence of biases seriously
mining the reality of the sequence, the arguments presented are not
conclusive. The more recent work devoted to this problem can be found
in Padovani et al. (2003) and Caccianiga \& Marcha (2003). The last
paper reports the detection of few cases of low-power blazar-like
sources, apparently not belonging to the spectral sequence. To exactly
evaluate the conclusions that one can derive from these studies, it is
important to note that the sequence has been constructed using
``typical'' blazars, characterized by jets well-aligned with the line
of sight. Beaming effects are rather sensitive to small misalignements
which translate in a substantial decrease of the luminosity but have a
smaller impact on the positions of the peaks (recall that the total
observed luminosity is proportional to $\delta ^4$, while the observed
peak frequencies $\propto \delta$). Therefore the sequence framework
{\it naturally predicts a large number of faint, slightly misaligned
jets}, with spectral properties similar to the well-aligned
counterparts.
A more direct and clear argument against the sequence would be the
discovery of blazars characterized by high peak frequencies (typical
of low-power BL Lacs), but with luminosity in the large-power range,
similar to FSRQs (Padovani et al. 2003). The discovery of such
sources, clearly falling in a ``forbidden'' region, would seriously
challenge the blazar sequence scenario.


The existence of the spectral sequence has been interpreted by
Ghisellini et al. (1998) (revisited in Ghisellini, Celotti \&
Costamante 2002) in terms of acceleration and cooling processes
determining the maximum energy of the emitting electrons. Reproducing
the SEDs by using a standard version of the synchrotron-IC model,
Ghisellini et al. (1998) discovered that the spectral sequence
translates into a rather tight anti-correlation existing between the
value of the energy $\gamma _b mc^2$ of the electrons emitting at the
peaks and the value of the total energy density (magnetic+radiation)
determining the total radiative cooling of the electrons
(Fig.\ref{donato}). This relation can be easily reproduced by assuming
that the value of $\gamma _b$ is the product of a balance between the
cooling rate (given by the amount of total energy density) and a
costant acceleration rate of the electrons. The most powerful sources
have a large amount of magnetic and radiation energy density,
determining a severe cooling and thus a small value for the
equilibrium Lorentz factor $\gamma _b$. On the contrary BL Lacs are
characterized by a low level of cooling, explaining the large Lorentz
factors of the electrons that acceleration is able to produce in these
sources.

A scenario that naturally accounts for most of the properties of
blazars is the so-called {\it internal shock scenario} (e.g. Spada et
al. 2001, see also Sikora \& Madejski 2001). It is assumed that the
relativistic flow is unsteady, with the engine ejecting a series of
``shells'' with different velocities. Faster shells will collide with
slower ones, forming a shock and dissipating a fraction of the total
kinetic energy energy (between 5\% and 10\%) into internal energy,
thought to be shared between magnetic field, electrons and protons. If
the flow has a typical Lorentz factor $\Gamma \sim 10-20$, with
dispersion $\Delta \Gamma / \Gamma \sim 1$ and the shells are ejected
with a typical timescale $t_{*}\sim r_g/c$ ($r_g$ is the gravitational
radius of the central black hole, with mass $M\sim 10^9 M_{\odot}$),
the distance at which most of the collisions happens is $d=ct_{*}
\Gamma ^2 \sim 10^{17}$ cm , close to the minimum distance necessary
to avoid absorption of $\gamma-$rays by soft photons produced by the
disk and BLR (Ghisellini \& Madau 1996). Initially developed to
account for the characteristics of high-power sources, the scenario
has been also applied to explain the low-power sources and to
reproduce the overall blazar sequence (Guetta et al. 2002).

\subsection{Jet power, composition, jet/accretion relation}

The possibility to infer the basic quantities associated to the flow,
offered by the modelling of the observed emission of blazars, gives us
a powerful tool to get insight into some of the most basic (and
important) problems related to the jet physics, such as the
composition (pair or proton dominated?), its power and eventually the
mechanisms involved in the launch of the jet and the link between
accretion and jet formation (e.g. Meier, Koide \& Uchida 2001). 

\begin{figure}[t]
\epsfxsize=15pc 
\epsfbox{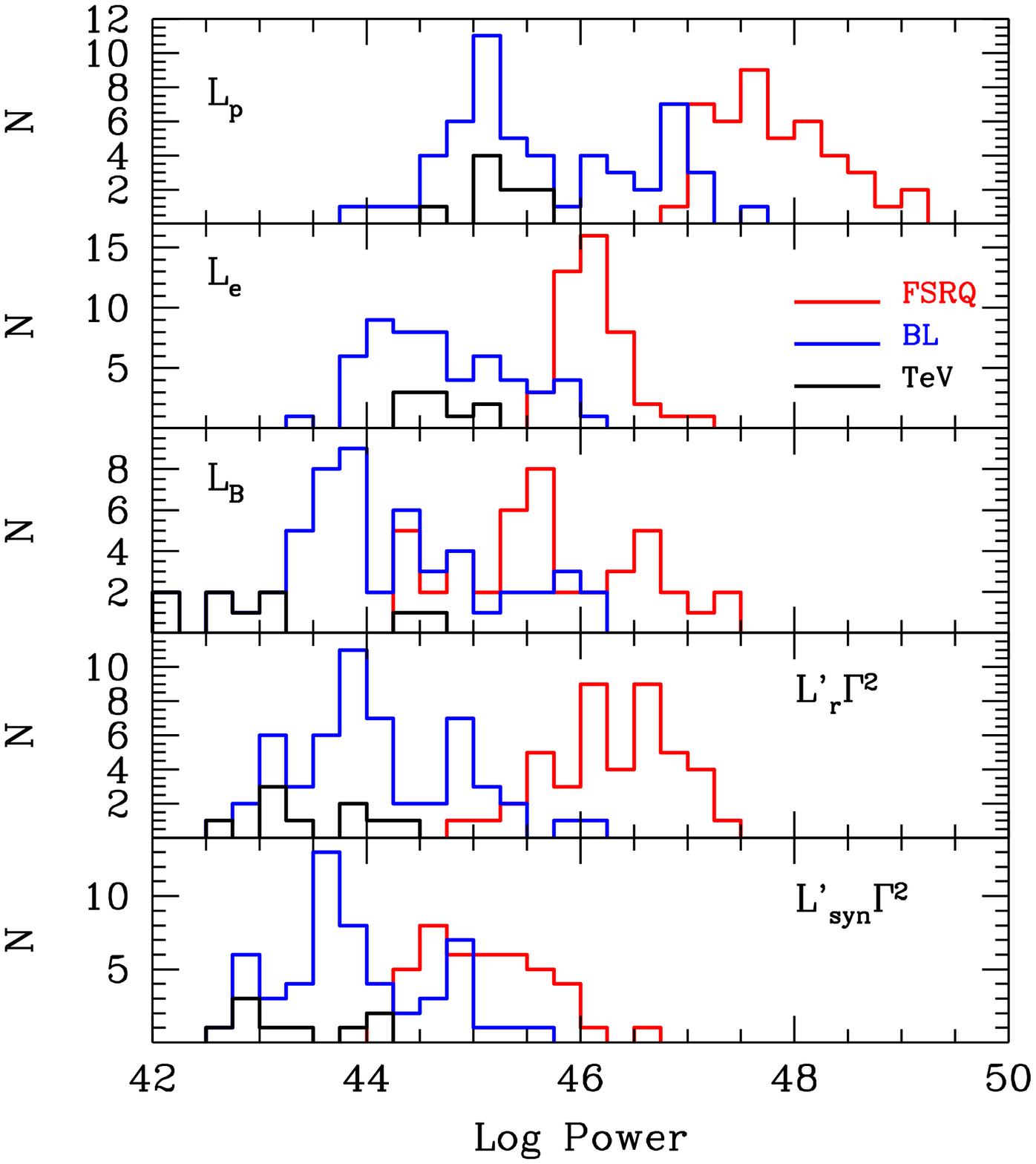}\epsfxsize=15pc\epsfbox{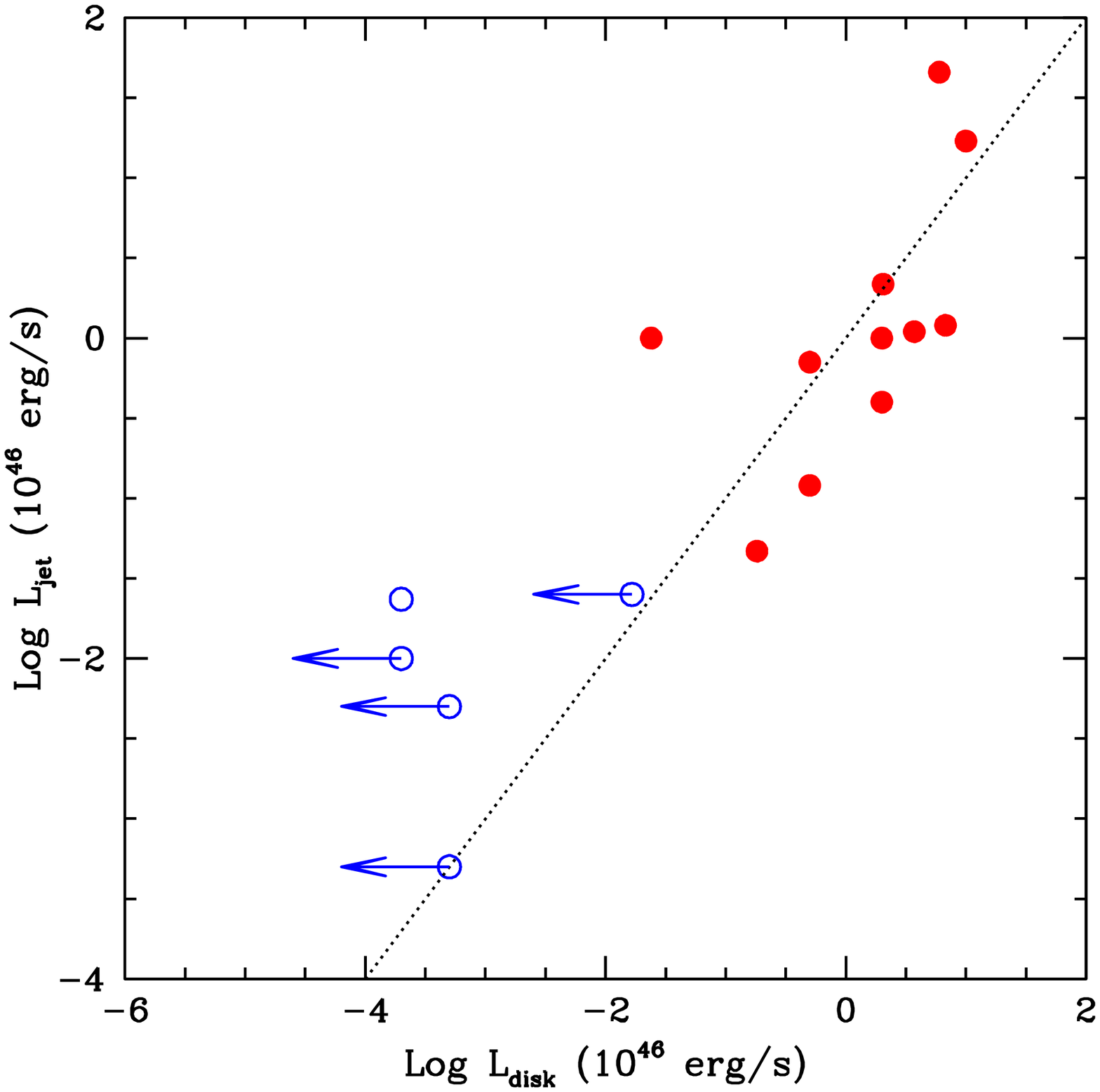} 
\caption{{\it Left:} Histograms of the kinetic powers separately
derived for (from top to bottom) protons, electrons, magnetic field
(from Ghisellini \& Celotti 2002). The fourth and the last panels
report the total radiative output and that of the synchrotron
component, respectively. Appears evident that electrons and magnetic
field only account for the emitted power, and therefore another
contribution (protons) must be admitted in order to permit the jet to
exit the galaxy and feed the lobes.  {\it Right:} Luminosities
released by the jet and the accretion flow derived for a sample of
blazars (from Maraschi \& Tavecchio 2003). In powerful sources the
amount of both emissions is similar, while in low-power BL Lacs the
accretion is underluminous with respect to the jet.
\label{ljet}}
\end{figure}

In Fig (\ref{ljet}) (from Ghisellini \& Celotti 2003) the power due to
electrons, magnetic field and protons (if present), estimated
modelling the SEDs of a sample of blazars with suitable
multiwavelength information, is compared to the radiative luminosity
of the jet. The basic condition that the flow must satisfy is that the
total power of the jet is larger than its radiative output, otherwise
the jet could not survive and reach the outer regions. This condition,
coupled to the estimates of the power transported by the electronic and
magnetic components, implies that the most important part of the power
is transported by some other components. A dinamically important
component of cold pairs can be excluded from the absence of the
Inverse Compton bump produced by these pairs, expected in the soft
X-rays (e.g. Sikora \& Madejski 2000).  Assuming that the jet is
composed by a ``normal'' plasma, with one (cold) proton per electron,
the total power comes out to be abot 10\% of the radiative
luminosity. Another possibility is that the source of the power is a
dominant magnetic field, with a value much larger than the field
inferres within the emitting region. Recently Blandford (2001)
proposed that the dominant role in powering jets is played by the
Poynting flux transported by a magnetic field, while the emission from
the jet just traces the ohmic dissipation of currents flowing from the
central black hole to the lobes and coming back to the nucleus.

Useful insights into the physical processes responsible for the
production of the jet can be obtained through the comparison of the
power shared by accretion and jet. In the past years several works
have ben devoted to the study of the connection between accretion and
the production of jets using available data (e.g. Rawling \& Saunders
1991, Celotti, Padovani \& Ghisellini 1997, Xu, Livio \& Baum 1999).
In the past most of these works were focussed on the determination of
the global parameters of jets at large scale (using the estimate of
the energy stored in radio-lobes) or at pc-scale (using VLBI data).
Using blazars, on the contrary, one can focus on the innermost regions
of jets, about two orders of magnitude smaller than accessible by
VLBI. In their important work, Rawling \& Saunders (1991) show that
the accretion luminosity and the jet power appear to be linearly
correlated. A strong correlation between the power released by
accretion and that channeled into the jet is naturally expected,
since, both in black-hole than in disk-powered models, the power
feeding the jet is related to the accretion rate (e.g. Livio, Ogilvie,
Pringle 1999). A linear relation between the two powers has been
inferred modelling the VLBI emission of a group of radio-sources by
Celotti, Padovani \& Ghisellini (1997), although the large scatter did
not allow the authors to draw firm conclusions. A work along these
lines but using blazars (both FSRQ and few BL Lacs) has been discussed
in Maraschi \& Tavecchio (2003). Fig. (\ref{ljet}) summarizes the
main results: for powerful sources jet and disk luminosities are
correlated and their values are comparable, $L_{\rm disk} \simeq
L_{\rm jet}$. This relation is no longer valid for BL Lacs sources,
for which the accretion flow appears underluminous with respect to the jet. A
possibility to reconcile this difference is that accretion in BL Lacs
has a lower efficiency than a standard disk (e.g. ADAF).

\subsection{TeV blazars: how high energy particles behave}

The family of TeV blazars is still a small but extremely interesting
class of sources. Until now six BL Lacs (Mkn 421, Mkn 501, PKS
2344+514, PKS 2155-304, 1ES1959+65, 1ES 1426+428) have a firm
detection.  Since the discovery of the first BL Lac object emitting
TeV radiation, Mkn 421, in 1992 (Punch et al. 1992) TeV blazars have
been the target of a very intense observational and theoretical
investigations.  Indeed the possibility to observed the emission
produced by very high energy electrons (up to Lorentz factors of the
order of $10^7$) coupled with observations in the X-ray band, where
the synchrotron peak of these sources is usually located, offers a
unique tool to probe the processes responsible for the acceleration of
relativistic particles. Investigations in the TeV band can also
provide important informations on the SED of the IR background
(e.g. Costamante et al. 2003 astro-ph/0308025). The spectral curvature
observed both in Mkn 501 and in Mkn 421 could be due to the absorption
of TeV photons by the extragalactic IR background (e.g. Aharonian et
al. 2002).


Studies conducted simultaneosly with X-rays and TeV are of particular
importance, since in the simple SSC framework one expects that
variations in X-rays and TeV should be closely correlated, being
produced by electrons with similar energies. Assuming a typical $B\sim
0.1$ G and $\delta\sim 10$, photons with energy E$\sim 1$ keV are
emitted by electrons with Lorentz factor $\gamma \sim 10^6$. The same
electrons will upscatter photons at energy $E\sim \gamma mc^2\sim 1$
TeV (since the Lorentz factor is extremely large the scattering will
occur in the KN regime even with optical target photons). In fact
observations at X-ray and TeV energies (Catanese et al. 1997, Buckley
et al. 1996) yield significant evidence of correlation between TeV and
X-rays. During the X/TeV 1998 campaign on Mkn 421 a flare has been
detected simultaneously both at X-ray and TeV energies and the
maxima were simultaneous within 1 hour, confirming that variations
into these two bands are closely related.  Subsequent more extensive
analisys confirmed these first evidences also in other sources. Note
however that the correlation seems to be violated in some cases, as
the observation of a ``orphan'' (i.e. not accompanied by the
correspondent X-ray flare) TeV event in the BLLac 1ES 1959+650
(Krawczynski et al. 2003) indicates (see also Horns 2002 for another
example of orphan flare detected in Mkn 421). If other examples of
such a behaviour will be found, the simplest versions of the SSC model
will be seriously in trouble. As discussed by Krawczynski et
al. (2003) a possible way out to the problem retaining the SSC model
is to admit a multi-component emission region, with a very dense
subregion responsible for the IC flare.  Other alternatives include a
contribution from external Compton, particular geometries for the
magnetic field, while hadronic models do not provide an explanation
for the particular flare of 1ES 1959+650.
 
Another possible problem for the SSC model (at least for the simplest
versions) is that the magnetic fields evaluated in the SEDs modelling
are quite below the equipartition value, with the particle energy
density dominating by 1-2 orders of magnitude over the magnetic one
(e.g. Kino et al. 2002). The most recent problem posed to SSC models
by observations is the large values of the Doppler factors inferred in
the modelling when the IR absorption of high-energy photons is
properly taken into account.

\subsection{The ``$\delta $ crisis''}

\begin{figure}[t]
\epsfxsize=25pc 
\centerline{\epsfbox{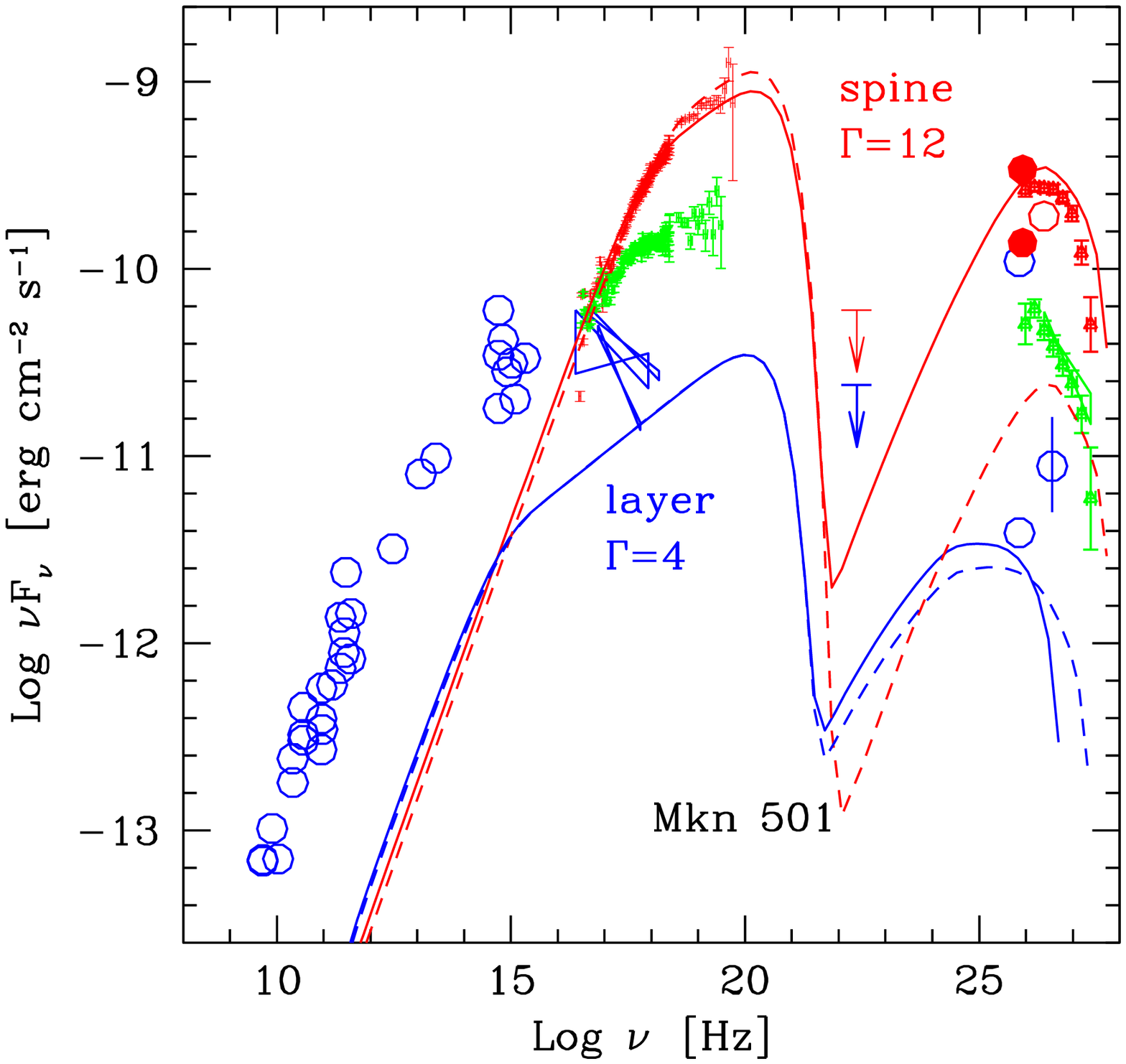}} 
\caption{The SED of Mkn 501 reproduced with a spine-layer model for
the jet (Ghisellini et al., in prep). A slower, external layer,
provides supplementary soft photons to the fast spine (and
vice-versa). Solid lines represent the {\it total} emission of the
spine and the layer, dashed lines represent the SSC emission
calculated taking into account only the synchrotron radiation produced
locally.
\label{spine}}
\end{figure}

In the recent past, different authors pointed out that, when IR
absorption is taken into account in the modelling of Mkn 421 and Mkn
501, rather large values of the Doppler factor, in the range 50-100,
are required (e.g. Krawczynski, Coppi \& Aharonian 2002, Konopelko et
al. 2003). This conclusion appears to be quite robust and independent
on the details of the model used to reproduce the data and the
expected IR background (see e.g. the discussion in Georganopoulos \&
Kazanas 2003).

On the other hand, there is growing evidence from VLBA studies
(Giroletti et al. 2003, Piner \& Edwards 2002, 2003) that jets at
pc-scale move slowly, since no superluminal components are visible in
the jet (although this evidence does not directly imply absence of
relativistic motion of the plasma in the jet). The largest
superluminal speed recorded in the TeV BL Lacs is about $4c$, measured
in PKS2155-304. Small speeds could be interpreted as due to a close
alignement of the jet with the line of sight (less than about 1
deg). However, such an almost perfect alignement is in contrast with
the typical angles needed by unification models: indeed (unless TeV BL
Lacs form a very special family) the number of parent FRI sources
inferred would be quite larger (more than one order of magnitude) than
the actual value.

These consideration lead to conclude that the jet must suffer a strong
deceleration from the blazar scale ($\sim $ 0.1 pc) to the VLBI ($\sim
$ 1 pc) scale. A large fraction of the power transported by the jet
should then be dissipated and probably radiated during this deceleration.
Georganopoulos \& Kazanas (2003) show that if a part of the dissipated
power is transformed into radiation the problem of the huge value of
$\delta $ can be partly overcome. These authors consider the emission
from a decelerating jet and note that the radiation from the outer
slowing-down portion of the jet can provide a supplementary soft
photon population for the IC scattering by electrons in the inner fast
jet. This emission is beamed into the fast jet frame and can easily
dominate over the local synchrotron radiation in producing high-energy
radiation.

An alternative scenario along the same lines (Ghisellini et al., in
prep) can be constructed assuming that the slowing down portion of the
jet is an external {\it layer}, decelerated by the interaction with
the external medium. The radiation produced into the layer (which can
be the site of intense particle acceleration, see e.g. Stawarz \&
Ostrowski 2002) enters into the fast {\it spine} and is used to
produce VHE photons through IC scattering.  This idea come from the
fact that there are several elements indicating that the jet in
low-power FRI radio-galaxies are composed by a fast moving internal
``spine'' and a low-moving outer layer. Interestingly also the images
of the VLBI scale jet in Mkn 501 shows such a structure, visible as a
limb-brightened structure of the jet (Giroletti et al. 2003).  As
discussed in the next section, in this framework relatively intense
high-energy emission is expected also from the layer, and this
emission can be visible from misaligned jets, i.e. radiogalaxies.

\section{Large scale jets}

\begin{figure}[t]
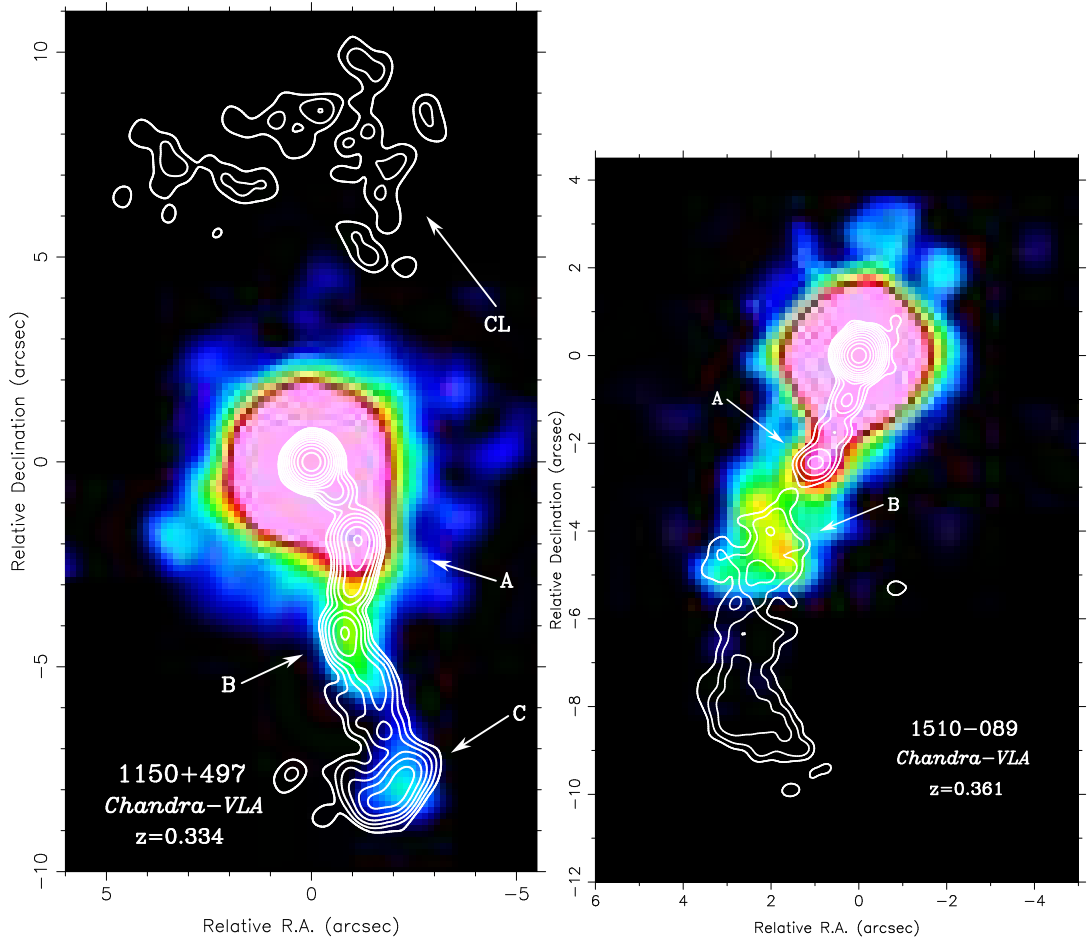

\epsfxsize=17pc 
\hspace*{-1.0 truecm }
\epsfbox{q1150xr.ps}\epsfxsize=17pc\epsfbox{q1510xr.ps} 
\caption{X-ray (colours) and radio (contours) maps of two jets studied
by Sambruna et al. (2004). PKS 1510-089 is also a well known
$\gamma-$ray blazar.
\label{chandra}}
\end{figure}

Although very common in the radio band, before the launch of the {\it
Chandra} satellite in 1999 only a handful of extragalactic kpc scale
jets were known to emit X-rays. Among them the bright and prominent
jets in 3C273 , M87, Cen A, studied with {\it EINSTEIN} and {\it
ROSAT}.  With the superior sensitivity and, especially, spatial
resolution of {\it Chandra} numerous jets have been detected,
triggering a new intense theoretical and observational work.  Even in
the source selected for the first light, the distant ($z=0.6$) quasar
PKS 0637-752, a prominent jet has been discovered (Chartas et
al. 2000; Schwartz et al. 2000).

Soon after the discovery of the X-ray jet in PKS 0637-0752 other jets
have been detected both in radio galaxies (both of the FRI and FRII
type) and in quasars. A recent census (see e.g. the WEB site maintened
by D. Harris and C. Cheung\footnote{\small
http://hea-www.harvard.edu/XJET/}) reported 25 jets detected in
X-rays. Different classes of AGNs are represented in the sample: the
most numerous group is that of radio-galaxies (both FRI and FRII), but
a large fraction is composed by powerful radio-loud QSOs. Most of the
jets have also an optical counterpart, in most cases detected by {\it
HST}.

The first problem posed by these observations is the identification of
the emission mechanism responsible for the production of X-rays. Since
these jets are known to emit in radio, the first candidate mechanism
is the extrapolation at high frequencies of the synchrotron
emission. In some cases (in particular for FRI jets) the the data are
consistent with a unique synchrotron component (steepening at high
frequencies probably because of radiatiove losses), but in other
several cases (especially in powerful quasars) this simple
interpretation fails to explain the observational evidence. This is
the case of the first jet discovered in PKS 0637-0752, for which two
separated emission components are clearly required by the shape of the
convex radio-optical-X-ray spectrum. The SSC model cannot explain the
level of the X-ray emission, since it would require very extreme
conditions. Tavecchio et al. (2000b) and Celotti, Ghisellini \&
Chiaberge (2000) proposed that the mechanism responsible for the
observed X-ray emission is the IC scattering of the CMB radiation by
relativistic electrons in the jet. The key point in this model is the
assumption that the jet in these (FRII) sources is still highly
relativistic (with bulk Lorentz factors $\Gamma =5-10$) at these
scales. It is also possible to reach the equipartition between
emitting electrons and magnetic field, and the kinetic power is
consistent with those usually derived for jets in powerful QSOs.

A systematic study of the emission from jets in QSOs has been
done using combined {\it Chandra-HST} observations by Sambruna et
al. (2004) (a first account is given in Sambruna et al. 2002).
We also started a program to image these jets uniformly at radio
wavelengths using the VLA, VLBA, and MERLIN.  Two examples of the jets
observed in X-rays and radio are reported in Fig. (\ref{chandra}) The
case of PKS 1510-089 is quite interesting, since this source is a
well-known blazar.  
The possibility to constrain the physical state of the plasma in the
jets both at blazars and kpc scale could offer us the interesting
opportunity to shed some light on the evolution of the jet from very
small scales, close to the central engine, to the outer regions, where
the jet is starting to significantly decelerate. In fact we are
starting to do this work for a small subset of the {\it Chandra} jets
for which good data for both regions are available (Tavecchio et al.,
submitted).  A first important result is that the powers independently
derived for both regions are very close, supporting the overall
metodology and suggesting the idea that the jet does not substantially
dissipate its power until its end. Another tempting element derives
considering the internal pressure of the knots derived with our
modelling. The derived values are in fact close to $10^{-11}$,
suggestively close to the pressure of typical hot gas halos found
around FRIs (Worrall et al. 2000).

Although the IC/CMB model appears to satisfactorily account for the
observative evidence, possible alternatives have been
presented. Dermer \& Atoyan (2002) show that the characteristics
convex shape of the radio to X-rays continuum can be reproduced by a
single synchrotron-emitting electron population continuously injected
and cooling through synchrotron and IC losses. While low-energy
electrons cool mainly through IC, high energy electrons cool mainly
through synchrotron, since KN effects severily limit Compton losses:
this effect translates into the convex shape of the spectrum. Another
possibility, involving emission from particle accelerated within the
boundary layer of the jet, has been discussed in Stawarz \& Ostrowski
(2002). Aharonian (2001) proposed that the high-energy emission comes
from the sychrotron emission by relativistic protons.

A problem faced by the IC/CMB model has been recently pointed out in
Tavecchio, Ghisellini \& Celotti (2003). In fact in this model, X-rays
are due to the emission by low-energy electrons ($\gamma \sim
100$). If the cooling of electrons is due to radiative losses the
lifetime of these electrons is virtually infinite. Therefore one
should expect that these electrons, streaming in the jet, would
produce a {\it continuous } emission, not concentrated in knots as
observed. Even assuming adiabatic losses due to the jet expansion the
result does not change. A possible solution is to admit that the
emission region is not uniform, but composed by several ``clumps'',
overpressured with respect to the environment: adiabatic losses
suffered by electrons due to the expansion of these small regions
should be enough to cool the particles.

Finally I note that the IC/CMB model seems to receive support from the
recent detection of X-ray emission from the jet in the high-redshift
($z=4.3$) quasar GB 1508+5714, characterized by an extreme X-ray/radio
flux ratio. As noted by Schwartz (2002), bright X-ray emission from
high-redshift sources due to IC/CMB is naturally predicted by the
increasing amount, proportional to $(1+z)^4$, of the energy density of
the CMB. Notably, the increasing importance of the IC/CMB emission with
the distance exactly counterbalances the decreasing surface brightness,
leading to speculate that IC/CMB emitting jets can be observed at
almost any redshift.

\subsection{Gamma-rays from large scale jets?}

An interesting consequence of all the models discussed above is that
large-scale jets should be sources of high-energy radiation at a level
possibly within the capabilities of the present instruments. Indeed
two FRIs have been indicated in the past as TeV emitters, namely
Centaurus A (Grindlay et al. 1975) and M87 (Aharonian et al. 2003b).

\begin{figure}[t]
\epsfxsize=25pc 
\centerline{\epsfbox{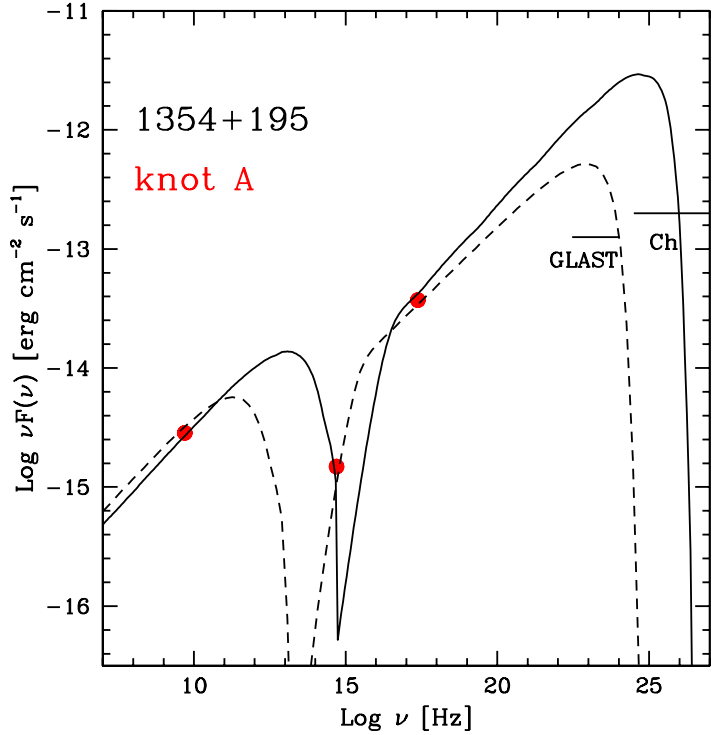}} 
\caption{Example of the predicted high-energy emission from large
scale jets. Radio, optical and X-ray data for a knot in the jet of the
QSO 1354+195 (from Sambruna et al. 2002) have been reproduced with
synchrotron-IC/CMB model. Solid and dashed line refer to the model
derived assuming that the optical is due to the synchrotron mechanism
or to the IC/CMB component, respectively. As discussed in the text,
the predicted gamma-ray fluxes is quite sensitive to this
assumption. The two horizonal lines report the sensitivity limit for
{\it GLAST} and the new generation of Cherenkov telescopes.
\label{1354}}
\end{figure}

The possibility that FRIs are TeV emitters, as BL Lacs (their) beamed
counterparts, has been discussed by Bai \& Lee (2001).  Stawarz et
al. (2003) pointed out that jets in FRIs must be natural candidates
for $\gamma -ray$ emission, since they contain high-energy electrons,
as indicated by the X-ray synchrotron emission. Analizing all the
possible sources for the IC scattering (synchrotron radiatiom,
emission from the inner blazar jet, radiation field of the galaxy)
Stawarz et al. (2003) conclude that we should expect rather bright
gamma-ray emission\footnote{Note, however, that TeV emission from
radiogalaxies could have a very different origin, as evidentiate
e.g. by Pfrommer \& Ensslin (2003), that proposed that gamma-rays are
produced in the elliptical host by hadron interaction of the cosmic
ray protons}. An alternative is that high-energy emission does not
originate into the large-scale jet, but it is produced in the small
scale jet (e.g. Protheroe, Donea \& Reimer 2003 for the specific case
of hadronic models). In this framework there is a possibility that the
spine-layer structure discussed above for TeV sources is a natural way
to produce bright high-energy emission from the layer (Ghisellini et
al., in prep).

Also powerful FRII jets in QSO should be good candidate for
high-energy emission. As an example in Fig. (\ref{1354}) I report the
SED of a knot in the jet of 1354+159, a source belonging to our {\it
Chandra-HST} survey. The IC/CMB model naturally predict a bright
emission at high energies, easily detecteble with the new
instrumentation. Note, however, that the predicted level of the high
energy flux (in particular in the Very High Energy band) critically
depends on the high energy limit of the emitting electrons. If the
optical emission belongs to the high energy tail of the synchrotron
emission (solid line) the predicted high-energy emission can easily
reach the TeV band. On the other hand if the optical comes from the
low-energy tail of the IC/CMB component emission at high energy is
much less conspicuous. Unfortunately the present data do not allow us
to discriminate between these two possibilities. Available multifilter
{\it HST} data suggest that the spectrum is quite steep, pointing
toward the synchrotron origin, but the large errorbars do not allow us
to draw a definitive conclusion. Another problem is related to the
fact the detection of these high-energy emission will be difficult,
due to the confusion with the (probably intense) emission of the
core. A possible way out to distinguish between small and large scale
origin could be the variability properties of the emission, since
radiation originating into the inner jet is rapide variable, while we
expect that the large scale emission forms a steady baseline.

\section{Concluding remarks}

From the previous pages it should come out how the study of the
emission from relativistic extragalactic jets and, in particular, of
their high energy emission, is an active and promising field, expected
to offer new important contribution to our global understanding of
relativistic outflows, which is quite far to be satisfactory.

In particular, the expected increased number of TeV blazars will allow
us to asses the importance of this emission for the general blazar
population, while the possibility to follow in great detail the time
evolution of the TeV emission in the brightest sources will continue
to stimulate time-dependent modelling of processes responsible for
acceleration and cooling of high-energy charged particles.

One of the surprises offered by {\it Chandra} has been the detection
of a large number of large-scale jets in the X-ray band. This evidence
suggests that large-scale jets could be sources of $\gamma $-rays
possibly detectable with the upcoming generation of high-energy
detectors.


\section*{Acknowledgments}
I am endebted with Laura Maraschi and Gabriele Ghisellini for the
continuous fruitful collaboration and stimulating interactions. I
thank Rossella Cerutti for the plot reported in Fig. 6. I am grateful
to Felix Aharonian who invited me to give this talk.

\end{document}